\documentclass[12pt]{article}
\usepackage{graphicx}
\makeatletter
\@addtoreset{equation}{section}
\makeatother

\let\la=\label  
  
 \def\bd{\begin{document}} \def\ed{\end{document}}
\def\ds{\documentstyle} \let\fr=\frac \let\bl=\bigl \let\br=\bigr
\let\Br=\Bigr \let\Bl=\Bigl
\let\bm=\bibitem
\let\na=\nabla
\let\pa=\partial \let\ov=\overline
\newcommand{\be}{\begin{equation}}
\newcommand{\ee}{\end{equation}}
\def\ba{\begin{array}}
\def\ea{\end{array}}
\newcommand{\ho}[1]{$\, ^{#1}$}
\newcommand{\hoch}[1]{$\, ^{#1}$}
\newcommand{\bea}{\begin{eqnarray}}
\newcommand{\eea}{\end{eqnarray}}
\newcommand{\ra}{\rightarrow}
\newcommand{\lra}{\longrightarrow}
\newcommand{\Lra}{\Leftrightarrow}
\newcommand{\ap}{\alpha^\prime}
\newcommand{\bp}{\tilde \beta^\prime}
\newcommand{\tr}{{\rm tr} }
\newcommand{\Tr}{{\rm Tr} }
\newcommand{\NP}{Nucl. Phys. }
\newcommand{\tamphys}{\it Michigan Center for Theoretical Physics,\\
Randall Laboratory, Department of Physics,
University of Michigan,\\ Ann Arbor, MI 48109-1120}

\newcommand{\auth}{M. J. Duff\footnote{Research supported in part by
DOE Grant DE-FG02-95ER40899 }}

\begin{document}
\begin{flushright}

\hfill\ \ \ {MCTP-04-39}\ \ \

{hep-th/0407175}\\
\end{flushright}

\hfill{}

\hfill{}

\vspace{24pt}

\begin{center}
\large{{\bf Benchmarks on the brane}}

\vspace{36pt}

\auth

\vspace{10pt}

{\tamphys}

\vspace{44pt}
\end{center}

 \abstract{ Branes now occupy center stage in theoretical physics
as microscopic components of M-theory, as the higher-dimensional
progenitors of black holes and as entire universes in their own right.
Their history has been a checkered one, however. Here we list some of the
milestones, starting with Dirac's 1962 paper.  Asim Barut was an early pioneer.}

\bigskip
\bigskip
\bigskip
\bigskip
\bigskip
\bigskip
\bigskip
\begin{center}
{\it ASIM BARUT MEMORIAL LECTURE, BOGAZICI UNIVERSITY, ISTANBUL,
OCTOBER 2002}
\end{center}
{\vfill\leftline{}\vfill}

\pagebreak
\setcounter{page}{1}

\tableofcontents
\newpage

{\it ``I know of no better way of teaching science than through its
history''}

Steven Weinberg

\section{Branes and Barut}

A perusal of the titles of papers appearing in the theory preprint archives
\cite{arXiv} reveals that {\it brane} now rivals {\it
string} as the noun of choice.  This is not without good reason.
Branes now occupy center stage in theoretical physics as microscopic
components of M-theory, as the higher-dimensional progenitors of black
holes and as entire universes in their own right.  However, while numerous
chronicles of string theory may be found in the literature, there is
comparatively little on branes. Their history has been a checkered one, having
been alternately revoked, reviled and revered.  Here we list some of
the milestones, starting with Dirac's 1962 paper.

This a particularly appropriate theme for this {\it Asim Barut Memorial Lecture}
since Asim was an early pioneer of brane theory
\cite{Barut:1973,Barut:1976,Barut:1988jh,Barut:ge,Barut:1992rq,Barut:zu,Barut:1992up,Barut:1994cw,Barut:vw}.
I first met Asim at the International Center for Theoretical Physics
in Trieste in 1972 where, like some other students of Abdus Salam, I
took my first postdoc.  Asim and I had frequent interactions
at the time and my memories are of (a) his continuing the great Turkish tradition
of group theory and its applications to quantum field theory and
particle physics, and (b) my playing along side him in the ICTP soccer
team, organized by Giuseppe Furlan.  Moreover, on a later visit to
Trieste in 1980 we discussed our common interest in the quantization of
p-forms, which was later to be of relevance for p-branes. He was an
obvious choice of speaker when Pope, Sezgin and I organized the 1989 Trieste
conference on supermembranes. It is an honor to be delivering this
year's Asim Barut Memorial Lecture.

\section{Nomenclature}

According to the Oxford English Dictionary \cite{OED}, first usage
of {\it brane} was by Duff, Inami, Pope, Sezgin and Stelle
in the May 1987 Trieste/CERN preprint published the following year in Nuclear
Physics B \cite{Inami}.  See Appendix \ref{OED1}.  The word had been invoked
earlier by Townsend in an unpublished lecture at the Trieste Spring
School, April 1987.  The lecture was intended to be entitled {\it
P-branes for pea-brains}, but organizer Ergin Sezgin baulked (at
pea-brains, not p-branes).  In this paper we shall always say {\it 5-brane}
rather than {\it fivebrane} and so on.

The names given to various branes have evolved as their place in the
scheme of things has become clearer.  For example, {\it M-theory} is a unified theory
involving branes that subsumes D=11 supergravity and the five D=10 superstring theories.
See Appendix \ref{OED2}. Following its discovery, the D=11 supermembrane and super 5-brane
became known as the {\it M2-brane} and {\it M5-brane} respectively.
(Discrete subgroups of) the Cremmer-Julia symmetries conjectured to be
dualities of the M2-brane became known as {\it U-dualities} of M-theory.
Similarly the Type II p-branes, first discovered as closed string
solitons carrying Ramond-Ramond charge, are now known as {\it
D-branes} following the realization by Polchinski \cite{Polchinski} that
they admit the dual open string interpretation of {\it
Dirichlet-branes} \cite{strings89,Dai}: surfaces of dimension p on which the
open strings can end\footnote{ ``No talk at Texas A\&M would be
complete without mention of supermembranes. If one compactifies the
Type I SO(32) superstring, which is oriented, and sends $r\rightarrow
0$, one obtains a theory with a super-D-brane...'' J. Polchinski,
Strings 89, Texas A\&M, March 1989 \cite{strings89}.}. At the same time
the heterotic and Type II
5-branes carrying Neveu-Schwarz charge were renamed {\it NS-branes}
and the fundamental string the {\it F-string} to distinguish it from
the D-string.  The D=6 dyonic string became known as the D1-D5-brane
system.  In charting the history of these various branes we shall adopt
the convention of using their modern names.  Moreover, we reserve the
name {\it D-brane} for the 1/2 BPS Type II branes whose mass equals their
charge and use the name {\it black branes} for those whose mass exceeds
their charge.

Reviews on branes may be found in
\cite{Duff:1987qa,Classical5,Khuristring,Duffsupermembranes5,Stelle5,
Polchinskistrings0,Dbranes}.
Reviews of $M$-theory may be found in
\cite{Schwarzpower,DuffM,TownsendM,Duffworld1,Kaku1,Kaku2,Ortin}.

\begin{figure}[h]\centering
\includegraphics[scale=0.5]{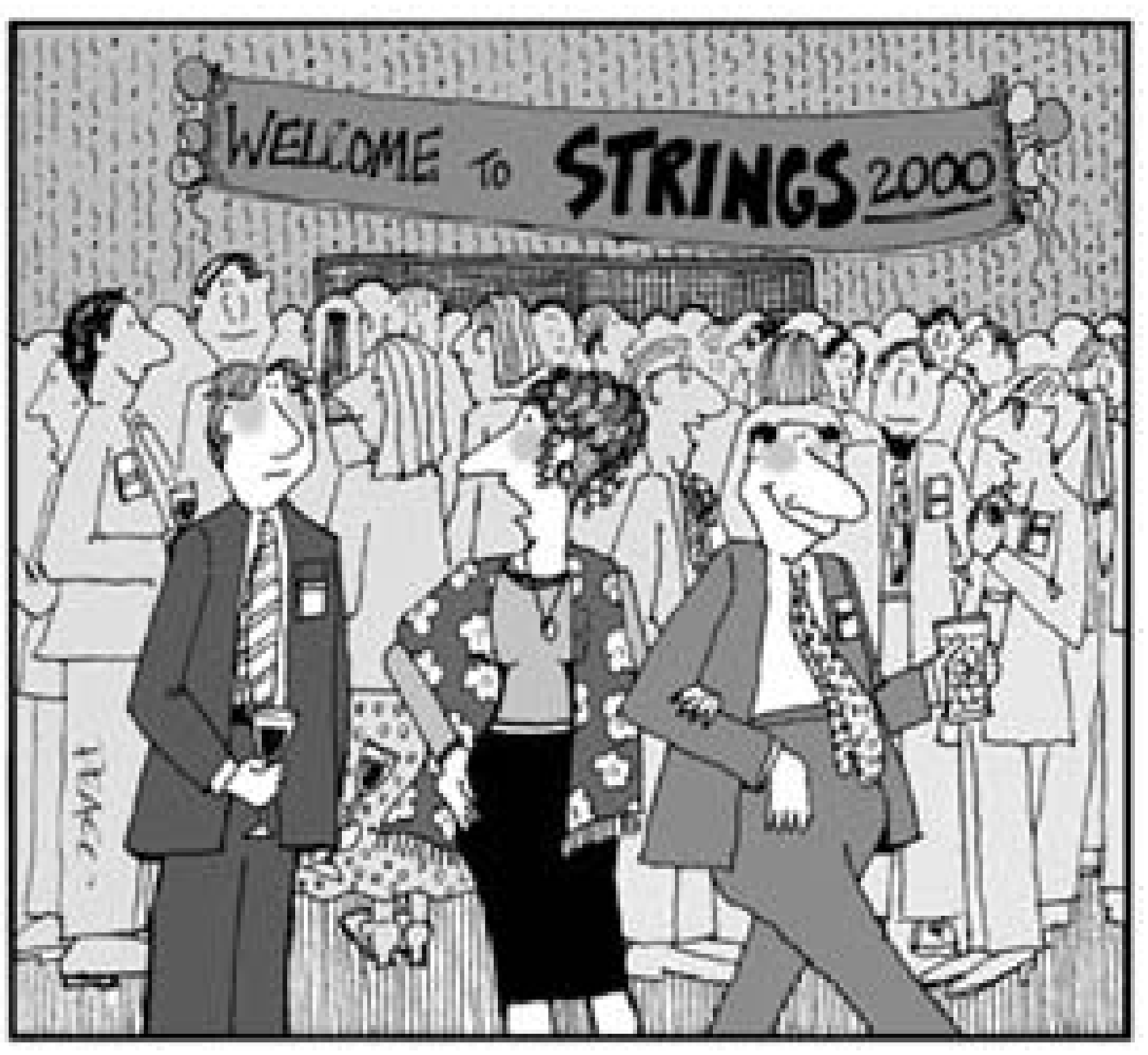}\newline
``Sorry. I know you've got charm but I'm really into
branes."\newline \tiny Copyright by Peaco Todd,
2000.\label{}\end{figure}

\section{What to include?}

Although, strictly speaking, strings are just 1-branes, we will not
here attempt a history of string theory since this has been covered
exhaustively elsewhere. For recent examples see
\cite{Wittenhistory,Veneziano,Schwarzparis}. We have,
however, included one or two milestones in string theory that have had a bearing
on branes. Similarly, we have included some discoveries in supersymmetric field
theories and supergravity that have proved influential in the development
of branes. In particular, the histories of Kaluza-Klein
compactifications, {\it dualities} and branes are all intertwined.

{\it Domain walls} have played an important part in the study of
topological defects and have sometimes been invoked in a cosmological
setting as the seeds of galaxy formation. They were first considered,
along with cosmic strings and magnetic monopoles, as solitons of flat-space
grand unified theories. Their development was thus separate from the
gravitational branes that are currently studied in M-theory, but there has
recently been a good deal of overlap and so they are included in the broad
definition of branes.

Our year-by-year summary is contained in the following sections which
have been divided onto four eras: (1) {\it Bosonic p-branes 1962-1986}, corresponding
to the years between Dirac's seminal paper and the discovery of
super p-branes; (2) {\it Super p-branes 1986-1995}, corresponding to
the period when the dominant view was that brane theory and eleven dimensions
represented a different approach to unification than ten-dimensional
superstrings; (3) {\it M-theory 1995-}, which followed the
discovery that both approaches are, in fact, part of a bigger
framework; (4) {\it Brane new world \footnote{This title is
borrowed from \cite{Castellani,Hawking:2000kj}, with apologies to William Shakespeare
and Aldous Huxley.} 2000-}, the present era of
brane and M-theory inspired attempts to understand the standard model
of particle physics, cosmology and black holes (I have not
attempted to detail these recent developments).  The citations quoted in a given year include not only the original
papers published in that year but also important subsequent
papers on the same topic that were published in later years.
Needless to say, the topics chosen reflect the bias of the author.

\section{Bosonic p-branes 1962-1986}
\la{timeline}

\noindent
{\bf 1962}

The Dirac action for a closed relativistic membrane as the 2+1
dimensional worldvolume swept out in spacetime \cite{Dirac}

\noindent {\bf 1971}

The Nambu-Goto action for a closed relativistic string
as the 1+1 dimensional worldsheet area swept out in spacetime
\cite{Nambu, Goto}

\noindent {\bf 1973}

Dual models for 3-branes based on the the conformal group $SU(2,2)$ \cite{Mansouri1, Mansouri2}

\noindent {\bf 1974}

Difficulties in extending dual models from strings to branes
\cite{Rivers,Kikkawa:1986dm}

\noindent {\bf 1976}

Discovery of supergravity \cite{Ferrara, Deser}

The role of domain walls in cosmology\cite{Kibble,Kibble2}

Mechanics of free relativistic bosonic branes \cite{Collins}

Local supersymmetry for spinning particles \cite{Brink:1976sz}

\noindent {\bf1977}

The Howe-Tucker action for a closed relativistic
membrane, with auxiliary metric \cite{Howetucker1}

S-duality (strong/weak coupling) in D=4 super gauge theories
\cite{Montonen,Witten:mh}

\noindent {\bf 1978}

Discovery of D=11 supergravity  \cite{Cremmerjuliascherk}

Spinning branes: branes with worldvolume supersymmetry
\cite{Howetucker2,Rocek}

A bag model involving a 3-form coupling \cite{Aurilia}

\noindent {\bf 1979}

Toroidal compactification of D=11 supergravity \cite{Cremmerjulia}

Cremmer-Julia hidden symmetries whose discrete subgroups were
subsequently known as U-dualities \cite{Cremmerjulia}

3-form potential of D=11 supergravity hints at a super 2-brane \cite{Julia:1979fw}
\newpage
\noindent {\bf 1980}

Supergravity vacua with 4-form flux and the cosmological constant
\cite{Duff:1980qv,Aurilia:1980xj}

Freund-Rubin vacua \cite{Freund:1980xh}

\noindent {\bf 1981}

The Polyakov action for a closed relativistic string, with
auxiliary metric \cite{Polyakov}

Dual formulation of $D=10$ supergravity with 6-form potential \cite{Chamseddine}

\noindent {\bf 1982}

Gauge-fixed branes and area-preserving diffeomorphisms
\cite{Goldstone,Hoppe1,Hoppe2}

Early braneworld proposal \cite{Akama} 

\noindent {\bf 1983}

Spacetime as a 3-brane in higher dimensions \cite{Rubakov}

Discovery of kappa symmetry for point particles \cite{Siegel}

K3 compactification of D=11 supergravity; moduli as scalar fields \cite{Duff:1983vj}

\noindent {\bf 1984}

The Green-Schwarz superstring: a string with spacetime supersymmetry \cite{Greenschwarz}

M-wave solution of D=11 supergravity \cite{Hull}

Anomaly cancellation in D=10 superstring theory \cite{Greenschwarz2}

T-duality for bosonic and heterotic strings
\cite{Kikkawa,Sakai,Nair,Sathiapalan}

\noindent {\bf 1985}

More spacetime as a 3-brane \cite{Pavsic:1984kv}

Dirac quantization rules for p-branes and their duals
\cite{Nepomechie,Teitelboim}

M-monopole solution of D=11 supergravity \cite{Han}

Discovery of the heterotic string \cite{Gross}

Calabi-Yau compactification of the heterotic string \cite{Candelas}

Spacetime as a 3-brane in a 5-dimensional bulk; trapping on the brane \cite{Visser}

\section {Super p-branes 1986-1995}

\noindent {\bf 1986}

Massive Type IIA supergravity \cite{Romans}

Kappa symmetry and the first super p-brane \cite{Hughes}

Branes as solitons \cite{Hughes}

\newpage

\noindent {\bf 1987}

The wordvolume action of the M2-brane \cite{Bergshoeff1,Bergshoeff2}

Type IIA superstring in D = 10 from the M2-brane in D = 11 \cite{Howe}

The brane-scan: a plot of spacetime dimension D versus brane worldvolume
dimension d allowed by supersymmetry (assuming scalar supermultiplets) \cite{Achucarro}

Massless spectrum of the M2-brane \cite{Bars1,Inami}

``The membrane at the end of the universe'': a CFT on the boundary of AdS
\cite{Bergshoeffsing,Duff:1987qa, Sutton,Bergshoeffgamma,Seibergwitten}

Brane quantization \cite{Fujikawa:1987av}

More spacetime as a 3-brane \cite{Gibbons:1986wg}

\noindent {\bf 1988}

{\it Brane} appears in print \cite{Inami}

Symmetries of bosonic branes \cite{Floratos:1988yp}

More branes as solitons \cite{Townsend4}

Heterotic NS 5-brane conjectured \cite{Duff:1987qa}

M2-branes and the cosmological constant \cite{Brown}

Matrix model of the M2-brane \cite{deWit1,deWit2}

Super p-branes and the signature of space-time \cite{Blencowe}

\noindent {\bf 1989}

Trieste conference on supermembranes \cite{Duffpopesezgin}

IIA/IIB, open/closed and oriented/unoriented string T-duality
\cite{Dine,Dai}

Dirichlet branes: surfaces on which open string end \cite{strings89,Dai,Leigh}

Supersymmetry algebra for extended objects \cite{Azcarraga}

\noindent {\bf 1990}

S-duality in four-dimensional string theory
\cite{Font:1990gx,Sen:1992fr,Senstrong}

NS-string solution of D=10 supergravity \cite{Dabholkar}

The heterotic gauge 5-brane solution of D=10 supergravity
\cite{Strominger1}

U-duality conjecture that the Cremmer-Julia hidden symmetries
of D=11 supergravity play the same role for M2-branes that S and T
duality play for strings  \cite{Luduality1,Luduality2,Luduality3}

Monopoles, strings and domain walls in D=4 from wrapped
5-brane \cite{Strominger1,Khuri,Liu,Khurinew}

\noindent {\bf 1991}

String/5-brane duality in D=10: the role of the 5-brane as the magnetic
dual of the string \cite{Luremarks,Lustrings,Luloop}

NS 5-brane solution of D=10 supergravity \cite{Luelem}

D-branes as closed string solitons \cite{Callan1,Callan2,
Luthree,Luscan,Khuristring}

Gauge fields on the D5-branes worldvolume  \cite{Callan1,Callan2,
Luthree,Luscan}

Black p-brane solutions of Type II supergravities \cite{Horowitz1,Gueven,Lublack,Dufflupopeblack}

The heterotic string as a soliton \cite{Lustrings,Harveystrominger,Hull:1995nu,Dabholkar:1995ep}

M2-brane solution of D=11 supergravity \cite{Duffstelle}

(${\cal N}=4,d=4$) gauge fields on the D3-brane worldvolume \cite{Luthree}

\noindent {\bf 1992}

The M5-brane solution of D=11 supergravity \cite{Gueven}

The new brane scan: allowing for vector (D-brane) and tensor
(M5-brane) worldvolume supermultiplets as well as scalar \cite{Luscan,
Khuristring}

D2-brane in D=10 from M2-brane in D=11, by dualizing a worldvolume
scalar into a vector \cite{Luscan}

\noindent {\bf 1993}

Black brane instabilities \cite{Gregory}

ADM mass formula for black branes \cite{Lu}

String/string duality in D=4:  S-duality for the fundamental string is
T-duality for a dual solitonic string \cite{Duffkhuri,Duffstrong}

The self-dual string solution of D=6 supergravity \cite{Lublack}

Dual heterotic string in D=6 from wrapping 5-brane on K3  \cite{Duffminasian}

AdS as the near-horizon geometry of p-branes
\cite{Gibbonstownsend,Gibbonsdufftownsend,Gibbons:1994vm}

\noindent {\bf 1994}

U-duality conjecture for Type II string: demonstration that elementary and
soliton states of (${\cal N}=8,D=4$) supergravity fall into U-duality multiplets,
with an explanation in terms of wrapped M2- and M5-branes
\cite{Hulltownsend}

IIA string compactified on K3 dual to heterotic string compactified on $T^{4}$
\cite{Hulltownsend,Comments}

\section {M-theory 1995-}
\label{M}

\noindent {\bf 1995}

Strong/weak coupling duality from the dual string \cite{Duffstrong}

D=11 Kaluza-Klein modes as extreme black holes of Type IIA string
\cite{Townsendeleven}

D=11 as strong coupling limit of Type IIA string
\cite{Townsendeleven,Wittenvarious}

Suggested D=11 unification of all five ten-dimensional superstrings,
with D=11 supergravity as its low-energy limit (subsequently known as
M-theory) \cite{Wittenvarious}

Heterotic and Type I strings related by strong/weak coupling duality
\cite{Wittenvarious,Hull:1995nu,Polchinskiwitten}

The anti-self dual (tensionless) string in D=6 \cite{Comments}

Black hole condensation, wrapped branes and topology change
\cite{Stromingermassless,Greene}

Dyonic string (also interpretable as D1-D5 system) solution of D=6 supergravity
\cite{Duff:1995yh}

M5-brane worldvolume anomaly \cite{Duffliuminasian,Wittenflux,Freed:1998tg}

D=11 origin of string/string duality:
a one-loop test \cite{Duffliuminasian}

M5 tension proportional to square of M2 tension
\cite{Duffliuminasian,Schwarzpower,Dealwis}

Role of branes in non-perturbative string theory \cite{Becker}

New p-brane solutions of $D \leq 9$ supergravity \cite{Lu:1995cs}

Dirichlet-branes and Ramond-Ramond charge \cite{Polchinski}

{\it M-theory} appears in print \cite{Schwarzpower,Horava1,Wittenfive}

Further ramifications of M-theory \cite{Schwarzpower,DuffM}

SU(N) gauge theories from bound state of N D-branes \cite{Wittenbound}

Heterotic string from M-theory on $S^{1}/Z_{2}$ \cite{Horava1,Horava2}

The D8-brane solution of massive Type IIA supergravity \cite{Bergshoeff4}

D-instanton and D7-brane solutions of Type IIB supergravity\cite{Gibbons:1995vg}

D-branes and topological field theories \cite{Bershadsky:1995qy}

All p-brane solutions of $4\leq D \leq 11$ supergravity \cite{Lu:1995yn}

Open p-branes \cite{Strominger3,TownsendD}

More D-branes from M-branes: D=11 derivation of D2 worldvolume action \cite{TownsendD}

Branes within branes \cite{Douglas}

5-branes and M-theory on an orbifold \cite{Wittenfive}

\noindent {\bf 1996}

Microscopic D1-D5 origin of the Bekenstein-Hawking entropy \cite{Stromingervafa}

Evidence for heterotic/heterotic duality \cite{DMW,Aspinwall,Berkooz}

D=6 heterotic phase transitions \cite{DMW,Seibergwittenphase,Duff:1996cf}

Orientifolds and D-branes \cite{Gimon}

F-theory \cite{Vafa}

Strong coupling of Calabi-Yau compactification \cite{Wittencalabi}

Intersecting branes \cite{Townsendpapa,Klebanov2,Behrndt2,Gauntlett}

Phase transitions in M-theory and F-theory \cite{Wittenphase}

BPS quantization of the 5-brane \cite{Dijkgraaf}

Classification of supergravity domain walls \cite{Cvetic:1996vr}

Branes with both worldvolume and spacetime supersymmetry
\cite{Howesezgin}

Worldvolume equations of the M5-brane \cite{Howesezgin,Howesezginp,
Aganagic}

D-branes and short distances in string theory \cite{Douglas:1996yp}

M-theory as a matrix model \cite{BanksM,Seibergwhy}

Worldvolume actions for D-branes \cite{Aganagic,Cederwall,Kappa}

Rotating brane solutions of D=11 supergravity \cite{Cvetic:1996ek}

\noindent {\bf 1997}

5-brane derivation of Seiberg-Witten theory \cite{Wittenfour}

Branes and the dynamics of QCD \cite{Wittenbranes}

D-branes and K-theory \cite{Moore,WittenD}

D-branes and the noncommutative torus \cite{Douglas:1997fm}

The AdS/CFT correspondence \cite{Maldacena,Gubser,Wittenads}

\noindent {\bf 1998}

Trieste conference on super 5-branes \cite{Duff:vs}

ADD large extra dimensions \cite{Arkani,Antoniadis:1998ig}

The brane anti-brane system and tachyon condensation\cite{Senstable,Sentachy}

\noindent {\bf 1999}

Randall-Sundrum: $AdS_{4}$, Minkowski or $dS_{4}$ spacetime as 3-brane in $AdS_{5}$
\cite{Randall1,Randall2,Karch}

D-branes and non-commutative geometry \cite{Seibergwitten2}

Brane-world black holes \cite{Chamblin:1999by}

Brane-world cosmology \cite{Flanagan:1999cu}

\section{Brane new world 2000-}

Numerous developments in the brane-world:

Brane-inspired particle phenomenology: attempts to embed the standard model
in M-theory via interesecting branes; strong, weak and electromagnetic
forces confined to a 3-brane in higher-dimensional bulk; insights into QCD
from the AdS/CFT correspondence.

Brane cosmology: the universe as a 3-brane; attempts to explain
inflation, dark energy and an accelerating universe; cosmic strings
attached to branes; the Big Bang as a brane-collision.

Braney black holes: deeper brane-inspired understanding of black hole
entropy and the information paradox, especially via the AdS/CFT
correspondence.

\section{Acknowledgements}

I would like to thank the Barut family for this invitation and Rami Gueven for his
kind hospitality during my visit to Istanbul.

\appendix
\section{Oxford English Dictionary: Brane}
\la{OED1}
\indent
 {\bf brane} , n.  Physics.

/Brit. breIn, U.S. breIn/[Shortened < membrane n.

I. Simple uses.

1. An extended object with any given number of dimensions, of which
strings in string theory are examples with one dimension. Also with
prefixed numbers, or symbols representing numbers, as 2-brane,
p-brane.

Quot. 1988 is from a paper received for publication earlier (18 May
1987) than quot. 1987 (1 Aug.).

{\bf 1987} Physics Lett. B. 198 441 The extension of the spacetime
supersymmetric Green-Schwarz covariant superstring action to
p-dimensional extended objects (p-branes) is possible if and only if
the on-shell p-dimensional Bose and Fermi degrees of freedom are
equal.

{\bf 1988} M. Duff et al. in Nucl. Physics B. 297 516 We shall be concerned
only with extended objects of one time and two space dimensions,
i.e. `2-branes'...  Possible `p-brane' theories exist whenever there
is a closed p + 2 form in superspace.

{\bf 1996} Sci. Amer. Jan. 75/2 He [sc. M. J. Duff] found that a
five-dimensional membrane, or a `five-brane', that moved through a
10-dimensional space could serve as an alternative description of
string theory.

{\bf 1997} New Scientist 18 Jan. 35/2 A string is a one-brane, an ordinary
membrane like a soap bubble is a two-brane, and so on.

{\bf 1998} Independent on Sunday 19 July (Sunday Rev.) 56/2 With extra
dimensions thrown in, strings turn into membranes..--and, to
complicate things further, these membranes are also called `p-branes',
where `p' is the number of dimensions.

{\bf 2000} Nature 2 Mar. 28/3 One of the key ideas..is that the
four-dimensional space-time we observe at everyday scales is actually
the evolution in time of a three-brane moving through an ambient
space-time of higher dimension.

II. Compounds.

2. { \bf brane-world}, a world model in which our space-time is the result of
a three-brane moving through a space-time of higher dimension, with
all interactions except gravity being confined to the three-brane.

{\bf 1999} Z. Kakushadze \& S.-H. H. Tye in Nucl. Physics B. 548 181 For
appropriate values of $V_{p-3}$ and $V_{p-9}$ all the known experimental
constants appear to be satisfied, and this scenario, which we will
refer to as `Brane World', a priori seems to be a viable possibility
for describing our universe.

{\bf 2001} Nature 28 June 987/3 Other variations of brane-world theory are
being used to tackle questions about the evolution of the Universe.

\section{Oxford English Dictionary: M-theory}
\la{OED2}
\indent
{\bf M-theory}, n.  Particle Physics.

Brit.  $\epsilon$m$\theta$ieri , U.S. $\epsilon$m$\theta$ieri, $\epsilon$m$\theta$eriÊ

[< M (app. representing MEMBRANE n.) + THEORY n.1]Ê

A unified theory involving branes that subsumes eleven-dimensional supergravity and the
five ten-dimensional superstring theories.

\smallskip
ÊÊ
Quot. 1996 is from a paper received for publication earlier (23 Oct. 1995) than quot.
1995 (17 Dec.).

{\bf 1995} Re: Confinement: Massive Gauge Bosons in sci.physics (Usenet newsgroup)
17 Dec., String theorists are a mathematically sophisticated crew, so I'm sure they would
enjoy an abstract description of the ÔM-theoryÕ (as it's called) from which one could
then derive all its varied guises.

{\bf 1996} J. H. Schwarz  in Physics Lett. B. 367 97/1
If one assumes the existence of a fundamental theory in eleven dimensions
(let's call it the ÔM theoryÕ), this provides a powerful heuristic basis for
understanding various results in string theory. [Note] This name was suggested by E. Witten.

{\bf 1998} Sci. Amer. Feb. 59/2 Despite all these successes, physicists are glimpsing only small
corners of M-theory; the big picture is still lacking.

{\bf 2002} U.S. News \& World Rep. 6 May
59/1 M-theory..holds that our universe may occupy just part of a many-dimensional
mega-universe. In that picture, it could be shadowed by another universe on a
different ÔbraneÕ- M-theory jargon for 3-D membrane.
Ê
\section{Where M stands for\ldots}

More M-etymology:

``Recent results indicate that if one assumes the existence of a
fundamental theory in eleven dimensions (let's call it the `M-theory'
[This name was suggested by E. Witten]), this provides a powerful
heuristic basis for understanding various results in string theory.''
J. Schwarz, hep-th/9510101.

``As it has been proposed that the eleven-dimensionl theory is a
supermembrane theory but there are some reasons to doubt that
interpretation, we will non-committally call it M-theory, leaving for
future the relation of {\it M} to membranes.'' P. Horava and E. Witten,
hep-th/9510209

``For instance, the eleven-dimensional `M-theory' (where M stands for
magic, mystery or membrane, according to taste) on $X \times S^{1}$,
with $X$ any ten-manifold, is equivalent to Type IIA on $X$, with a
Type IIA string coupling constant that becomes small when the radius
of $S^{1}$ goes to zero.'' E. Witten, hep-th/9512219

\end{document}